\documentclass[onecolumn]{aa}  
\usepackage{natbib}
\usepackage{twoopt}
\usepackage{ wasysym }
\usepackage{txfonts}
\usepackage{color}
\begin{document} 

\title{Neutron star matter equation of state including  $d^*$-hexaquark degrees of freedom}

\author{ A. Mantziris \inst{1,2}  \and A. Pastore \inst{1} \and I. Vida\~na \inst{3} \and D. P. Watts \inst{1} \and M. Bashkanov \inst{1} \and A. M. Romero \inst{1}}

\institute{ Department of Physics, University of York, Heslington, York, Y010 5DD, United Kingdom \and  Department of Physics, Imperial College London, London SW7 2AZ, United Kingdom  \and  INFN Sezione di Catania, Dipartimento di Fisica ``Ettore Majorana", Universit\`a di Catania, Via Santa Sofia 64, I-95123 Catania, Italy}

\date{\today}

%%%%%%%%%%%%%%%%%%%%%%%%%%%%%%%%%%%%%%%%%%%%%%%%%%%%%%%%%%%%%%%%%%%%%%%%%%%%%%%%%%%%%%%%%%%%%%

\abstract{
We present an extension of a previous work where, assuming a simple free bosonic gas supplemented with a relativistic meand field model to describe the pure nucleonic part of the EoS, we studied the consequences that the first non-trivial hexaquark $d^*$(2380) could have on the properties of neutron stars.
Compared to that exploratory work we employ a standard non-linear Walecka model including additional terms that describe the interaction of the $d^*(2380)$ di-baryon with the other particles of the system through the exchange of $\sigma$- and $\omega$-meson fields. 
Our results have show that the presence of the $d^*(2380)$ leads to maximum masses compatible with the recent observations of $\sim 2$M$_\odot$ millisecond pulsars if the interaction of the $d^*(2380)$ is slightly repulsive or the $d^*(2380)$ does not interacts at all. An attractive interaction makes the equation of state too soft to be able to support a $2$M$_\odot$ neutron star whereas an extremely repulsive one induces the collapse of the neutron star into a black hole as soon as the $d^*(2380)$ appears.
}


\maketitle

%%%%%%%%%%%%%%%%%%%%%%%%%%%%%%%%%%%%%%%%%%%%%%%%%%%%%%%%%%%%%%%%%%%%%%%%%%%%%%%%%%%%%%%%%%%%%%

\section{Introduction}

Neutron stars are the remnants of the gravitational collapse of massive stars during a supernova event of Type-II, Ib or Ic. Their masses and radii are typically of the order of $1-2 M_\odot$  ($M_\odot \simeq 2 \times 10^{33}$g is the mass of the Sun) and $10-14$ km, respectively. With central densities in the range of $4-8$ times the normal nuclear matter saturation density, $\epsilon_0 \sim 2.7 \times 10^{14}$ g/cm$^3$ ($\rho_0 \sim 0.16$ fm$^{-3}$), neutron stars are most likely among the densest objects in the Universe (\cite{shapiro,glendenning,haensel,pharos}). These objects are, therefore, excellent laboratories to test our present understanding of the theory of strong interacting matter at extreme conditions, offering an interesting interplay between the physics of dense matter and astrophysical observables.

The conditions of matter inside neutron stars are very different from those encountered on Earth. A good theoretical knowledge of the nuclear equation of state (EoS) of dense matter is, therefore, required to understand the properties
of neutron stars. Its determination, however, is very challenging due to the wide range of densities, temperatures and isospin asymmetries found in these objects, and it constitutes nowadays one of the main problems in nuclear astrophysics. 
The main difficulties are associated to our lack of a precise knowledge of the behavior of the in-medium nuclear interaction, and to the very complicated resolution of the so-called nuclear many body problem (\cite{mbp}). 

The nuclear EoS has been largely studied by many authors using both phenomenological and microscopic many-body approaches. Phenomenological ones, either non-relativistic or relativistic, are based on effective interactions that are frequently built to reproduce the properties of nuclei (\cite{stone}). Skyrme interactions (\cite{skyrme59,skyrmea,davesne,grasso}) and relativistic mean-field (RMF) models (\cite{rmfbb,rmfa,rmfb}) are among the most used ones. Many of such interactions are built to describe nuclear systems close to the isospin symmetric case and, therefore, predictions at high isospin asymmetries should be taken with care. Most Skyrme interactions are, by construction, well behaved close to  $\rho_0$ and moderate values of the isospin asymmetry. However,  only certain combinations of the parameters of these forces are well determined experimentally. As a consequence, there exists a large proliferation of different Skyrme interactions that produce a similar EoS for symmetric nuclear matter, but predict a very different one for pure neutron matter. Few years ago, Stone {\it et al.} (\cite{stoneb}) made an extensive and systematical test of the capabilities of several existing Skyrme interactions to provide good neutron star candidates, finding that only few of these forces passed the restrictive tests imposed.
%A more stringent constraint has been recenty done by Dutra {\it et al.} (\cite{dutra}) who have examined the suitability of 240 Skyrme interactons with respect to 11 constraints derived from experimental data and the empitical properties of symmetric matter at and close to saturation. These authors found that only 5 of the 240 analyzed satisfied all the constraints imposed. 

Relativistic mean-field models are based on effective Lagrangians densities where the interaction between baryons is described in terms of meson exchanges. The couplings of nucleons with mesons are usually fixed by fitting masses and radii of nuclei and the properties of nuclear bulk matter, whereas those of other baryons, like hyperons, are fixed by symmetry relations and hypernuclear observables. Recently, Dutra {\it et al.,} (\cite{dutrab}) have analyzed, as in the case of Skyrme, several parametrizations of 7 different types of  RMF models imposing constraints from symmetric nuclear matter, pure neutron matter, symmetry energy and its derivatives finding that only a very small number of these parametrizations is consistent with all the nuclear constraints considered in that work.

Microscopic approaches, on other hand, are based on realistic two- and three-body forces that describe scattering data in free space and the properties of the deuteron. These interactions are based on meson-exchange (\cite{m1,m2,m3,m4,m5,m6,m7,m8,m9,m10}) or, very recently, on chiral perturbation theory (\cite{xft1,xft2,xft3,xft4}). To obtain the nuclear EoS one has to solve then the complicated many-body problem whose main difficulty lies in the treatment of the repulsive core, which dominates the short-range of the interaction. Different microscopic many-body approaches has been extensively used for the study of the nuclear matter EoS. These include among others: the Brueckner--Bethe--Goldstone (\cite{mbp,bbg}) and the Dirac--Brueckner--Hartree--Fock (\cite{dbhf1,dbhf2,dbhf3}) theories, the variational method (\cite{var}), the correlated basis function formalism (\cite{cbf}), the self-consistent Green's function technique (\cite{scgf1,scgf2}) or the $V_{\mbox{low}\,k}$ approach (\cite{vlowk}). The interested reader is referred to any of the quoted works for details on these approaches.

Nowadays, it is still an open question which is the true nature of neutron stars. Traditionally the core of neutron stars has been modelled as a uniform fluid of neutron-rich nuclear matter in equilibrium with respect to the weak interaction ($\beta$-stable matter). Nevertheless, due to the large value of the density, new  degrees of freedom are expected to appear in addition to nucleons. Examples of these new degrees of freedom widely studied include pion (\cite{pion}) and kaon (\cite{kaon}) condensates, hyperons (\cite{vidanaepja,vidanaprs}), $\Delta$ isobars (\cite{delta1,delta2,delta3}), deconfined quarks (\cite{quark}) or even di-baryonic matter (\cite{dibaryon}).  The most precise and stringent neutron star constraint on the nuclear EoS comes from the recent determination of the unusually high masses of the millisecond pulsars PSR J1614-2230 (\cite{demorest}), PSR J0348+0432 (\cite{antoniadis}) and PSR J0740+6620 (\cite{cromartie}). These three measurements imply that any reliable model for the nuclear EoS should predict maximum masses at least larger than $2 M_\odot$. This observational constraint rules out many of the existent EoS models with exotic degrees of freedom, although their presence in the neutron star interior is, however, energetically favorable. This has lead to puzzles like the \textquotedblleft hyperon puzzle\textquotedblright (\cite{vidanaepja}) or the \textquotedblleft$\Delta$\textquotedblright puzzle (\cite{delta1,delta2}) whose solutions are not easy and presently are subject of very active research.

Recently, we studied the role of a new degree of freedom $d^*(2380)$ (\cite{bashkanov2019deuteron}) on the nuclear EoS (\cite{vid18}). The $d^*(2380)$ is a massive positively charged non-strange particle with integer spin (J=3) and it represents the first known non-trivial hexaquark evidenced in experiment (\cite{MB1,MBC,MBE1}). The importance of such a new degree-of-freedom resides in the fact that it has the same  $u, d$  quark composition as neutrons and protons and, therefore, does not involve any strangeness degrees of freedom. Moreover, it is a boson and as such it may condensate within the star. In our previous work we showed that despite its very large mass, the $d^*(2380)$ can appear in the neutron star interior at densities similar to those  predicted for the appearance of other nucleon resonances, such as the $\Delta$, or hyperons. That work was a first attempt to study the consequences that the presence of the $d^*(2380)$ could have on the properties of neutron stars where, however, we assumed the $d^*(2380)$ as simple gas of non-interacting bosons. We have, therefore, decided to pursue a more detailed study which accounts for explicit interaction of $d^*(2380)$ with the surrounding medium. To this aim, we employ a standard non-linear Walecka model (\cite{dutrab}), within the framework of a relativistic mean field theory (RMF). Starting from a well-known nucleonic Lagrangian (\cite{glendenning,delta1}), we employ the established $d^*(2380)$ properties to determine its interaction with other particles. In particular, we aim at providing first constraints on the sign (attractive or repulsive) for the effective interaction of such a particle.

The manuscript is organized in the following way. The Lagrangian density including the $d^*(2380)$ is shortly presented in Section \ref{Sec:Sky}. Our main results regarding the appearance and effect of $d^*(2380)$ on neutron stars are shown and discussed in Section \ref{sec:res}. Finally, our concluding remarks and possible directions for future work  are given in Section \ref{sec:concl}.

%%%%%%%%%%%%%%%%%%%%%%%%%%%%%%%%%%%%%%%%%%%%%%%%%%%%%%%%%%%%%%%%%%%%%%%%%%%%%%%%%%%%%%%%%%%%%%

\section{Lagrangian density}
\label{Sec:Sky}

The total Lagrangian density of a system that is composed of nucleons ($N=n,p$), the four $\Delta$ isobar resonances ($\Delta=\Delta^-,\Delta^0,\Delta^+,\Delta^{++}$), leptons ($l=e^-,\mu^-$), scalar-isoscalar ($\sigma$), vector-isoscalar ($\omega$) and vector-isovector ($\rho$) mesons, and includes in addition the $d^*(2380)$ di-baryon is simply given by    
\begin{equation}
\mathcal{L}=\sum_{N} \mathcal{L}_N+\sum_{\Delta} \mathcal{L}_\Delta+\sum_{l} \mathcal{L}_l+\mathcal{L}_m+\mathcal{L}_{d^*} \ ,
\label{eq:lagden}
\end{equation}
where
\begin{eqnarray}
\mathcal{L}_N&=&\bar{\Psi}_N\left[ i\gamma_\mu\partial^\mu-m_N+g_{\sigma N}\sigma -g_{\omega N}\gamma_\mu\omega^\mu-g_{\rho N}\gamma_\mu\frac{\boldsymbol{\tau_N} \cdot \boldsymbol{\rho}^\mu}{2}\right]\Psi_N \ , \nonumber \\
\mathcal{L}_\Delta&=&\bar{\Psi}_{\Delta\nu}\left[ i\gamma_\mu\partial^\mu-m_\Delta+g_{\sigma \Delta}\sigma -g_{\omega \Delta}\gamma_\mu\omega^\mu-g_{\rho \Delta}\gamma_\mu \boldsymbol{I}_\Delta\cdot\boldsymbol{\rho}^\mu\right]\Psi_\Delta^\nu \ , \nonumber \\
\mathcal{L}_l&=&\bar{\Psi}_l\left[ i\gamma_\mu\partial^\mu-m_l\right]\Psi_l \ , \nonumber \\
\mathcal{L}_m&=& \frac{1}{2}\partial_\mu \sigma \partial^\mu\sigma-\frac{1}{2}m_\sigma^2 \sigma^2-\frac{1}{3}b m_N g_{\sigma N}^3 \sigma^3-\frac{1}{4}c g_{\sigma N}^4 \sigma^4+\frac{1}{2}m_\omega^2\omega_\mu\omega^\mu
-\frac{1}{4}\Omega_{\mu\nu} \Omega^{\mu\nu} -\frac{1}{4}\mathbf{R}_{\mu\nu}\mathbf{R}^{\mu\nu}+\frac{1}{2}m_\rho^2 \boldsymbol{\rho}_\mu \cdot \boldsymbol{\rho}^\mu \nonumber \\
\mathcal{L}_{d^*}&=&(\partial_\mu-ig_{\omega d^*} \omega_\mu)\phi^*_{d^*}(\partial^\mu+ig_{\omega d^*} \omega^\mu)\phi_{d^*}-(m_{d^*}-g_{\sigma d^*}\sigma)^2\phi^*_{d^*}\phi_{d^*} \ .
\end{eqnarray}
\noindent $\Psi_N$ and $\Psi_\Delta^\nu$ are the Dirac and Rarita--Schwinger fields for the nucleon and the $\Delta$ isobar, respectively;  $\Psi_l$ is the lepton Dirac field, $g$ represents the different baryon-meson couplings;  and $\boldsymbol{\tau_N}$ and $\boldsymbol{I}_\Delta$ are isospin 1/2 and 3/2 operators. The strength tensor of the $\omega$ and $\rho$ mesons is denoted by $\Omega_{\mu\nu}=\partial_\mu\omega_\nu-\partial_\nu\omega_\mu$ and $\boldsymbol{R}_{\mu\nu}=\partial_\mu\boldsymbol{\rho}_\nu-\partial_\nu\boldsymbol{\rho}_\mu-g_{\rho N}\left(\boldsymbol{\rho}_\mu\times\boldsymbol{\rho}_\nu\right)$ whereas the parameters $b$ and $c$, associated with the non-linear self-interactions of the $\sigma$ field, guarantee that the value of the incompressibility of nuclear matter is within the experimental range. Finally, $\phi_{d^*}$ indicates the wave function of the $d^*(2380)$ condensate. The masses of the nucleons, $\Delta$ isobars and leptons are denoted by $m_N, m_\Delta$ and $m_l$. The main properties of the nucleons, $\Delta$'s and the $d^*(2380)$ di-baryon are summarized in Tab.\ref{particle_table}.  

%%%%%%%%%%%%
\begin{table}[t]
\begin{center}
	\scalebox{1.0}{
	\begin{tabular}{|c|c|c|c|c|c|c|}
	\hline
	    &$m$ (MeV) & $J$ &$I$ & $I_3$ & $b$ & $q$    \\ \hline
	$n$ &939 &1/2 &1/2 & -1/2 &1  & 0  \\ \hline
	$p$ &939 &1/2 &1/2 &1/2  &1 & 1  \\ \hline
	$\Delta^-$ &1232 &3/2 &3/2 &-3/2&1 &-1    \\ \hline
	$\Delta^0$ &1232 &3/2 &3/2 &-1/2&1 &0    \\ \hline
	$\Delta^+$ &1232 &3/2 &3/2 &1/2&1 &1    \\ \hline
	$\Delta^{++}$ &1232 &3/2 &3/2 &3/2&1 &2    \\ \hline
	$d$* &2380 &3 &0 &0&2 &1   \\ \hline
	\end{tabular}}
\end{center}
	\caption{Mass ($m$), spin ($J$), isospin ($I$), isospin third component ($I_3$), baryon number ($b$) and electric charge ($q$) of nucleons, $\Delta$'s and the di-baryon $d^*(2380)$.}
	\label{particle_table}
\end{table}
%%%%%%%%%%%%

In a RMF description of infinite nuclear matter, the meson fields are treated as classical fields. Meson field equations in the mean field approximation can be easily derived by applying the Euler--Lagrange equations to the Lagrangian density of Eq.\ (\ref{eq:lagden}) and
replacing field operators by their ground-state expectation values $\sigma\rightarrow\bar{\sigma},\,\,\omega_\mu\rightarrow \omega_0,\,\,\text{\boldmath{$\rho$}}_\mu\rightarrow \bar{\rho}_0^{(3)}$. They read simply
\begin{eqnarray}
m^2_\sigma \bar{\sigma}&=&\sum_{B=N,\Delta} g_{\sigma B}\left( \frac{2J_B+1}{2\pi^2}\int_0^{k_{F_B}}\frac{m^*_B}{\sqrt{k^2+m^{*2}_B}}k^2dk\right)-b m_N g_{\sigma N}\left( g_{\sigma N}\bar{\sigma}\right)^2-c g_{\sigma N}\left( g_{\sigma N}\bar{\sigma}\right)^3+ g_{\sigma d^*}\rho_{d^*} \\ 
m^2_{\omega} \bar{\omega}_0 &=&\sum_{B=N,\Delta} g_{\omega B} \frac{2J_B+1}{6\pi^2}k_{F_B}^3 -g_{\omega d^*}\rho_{d^*} \\
m^2_{\rho}\bar{\rho}_0^{(3)}&=&\sum_{B=N,\Delta} g_{\rho B} \frac{2J_B+1}{6\pi^2}k_{F_B}^3 I_{3B} \ ,
\end{eqnarray}
where $J_B$ is the spin of the baryon B, $k_{F_B}$ it Fermi momentum, $m^*_B=m_B-g_{\sigma B}$ its effective mass, $I_{3B}$ the third component of its isospin, and $\rho_{d^*}=2m_{d^*}\phi^*_{d^*}\phi_{d^*}$ is the density of the $d^*(2380)$ di-baryon. The energy density and the pressure of the system are  obtained from the energy-momentum tensor
\begin{eqnarray}
\varepsilon&=&\sum_{B=N,\Delta} \frac{(2J_B +1)}{2\pi^2}\int_0^{k_{F_B}}\sqrt{k^2+m^{*2}_B}k^2 dk+\sum_{l=e^-,\,\,\mu^-}\frac{1}{\pi^2}\int_0^{k_F^l}\sqrt{k^2+m^2_l}k^2 dk+\frac{1}{2}m_\sigma^2 \bar{\sigma}^2+\frac{1}{3} b m_N (g_\sigma \bar{\sigma})^3+\frac{1}{4}c(g_\sigma 
\bar{\sigma})^4\nonumber\\
&+&\frac{1}{2}m_\omega^2 \bar{\omega}_0^2+\frac{1}{2}m_\rho^2(\bar{\rho}_0^{(3)})^2+m^*_{d^*}\rho_{d^*}\label{eq:energy}\\
P&=&\sum_{B=N,\Delta} \frac{(2J_B+1)}{6\pi^2}\int_0^{k_{F_B}}\frac{k^4 dk}{\sqrt{k^2+m^{*2}_B}} +\sum_{l=e^-,\,\,\mu^-}\frac{1}{3\pi^2}\int_0^{k_F^l}\frac{k^4 dk}{\sqrt{k^2+m^2_l}}-\frac{1}{2}m_\sigma^2 \bar{\sigma}^2-\frac{1}{3} b m_N (g_\sigma \bar{\sigma})^3 - \frac{1}{4}c(g_\sigma 
\bar{\sigma})^4\nonumber\\
&+&\frac{1}{2}m_\omega^2 \bar{\omega}_0^2+\frac{1}{2}m_\rho^2(\bar{\rho}_0^{(3)})^2 \ .
\label{eq:press}
\end{eqnarray}
Note that the $d^*(2380)$ does not contribute to the total pressure of the system. Chemical equilibrium in the neutron star interior without neurtrino trapping leads to the following relations between the chemical potentials of the different species:
\begin{equation}
\mu_i=b_i\mu_n - q_i\mu_e\;,
\end{equation}
where $b_i$ and $q_i$ are the baryon number and the electric charge of the particle {\it i}. The chemical potentials of the different particle species read
\begin{eqnarray}
\mu_B&=&\sqrt{k_{F_B}^2+m_B^{*2}}+g_{\omega B}\bar{\omega}_0 +g_{\rho B} I_{3B} \bar{\rho}_0^{(3)} \ , \,\,\, (B=n,p,\Delta^-, \Delta^0, \Delta^+, \Delta^{++}) \;, \\
\mu_{d^*}&=&m_{d^*}-g_{\sigma d^*}\bar{\sigma}-g_{\omega d^*}\bar{\omega}_0\;,\\
\mu_l&=&\sqrt{k_{F_l}^2+m_l^2} \ , \,\,\, (l=e^-,\mu^-)\;.
\end{eqnarray}

Let us finish this section with a short discussion of our choice of the different coupling constants. The nucleon couplings $g_{\sigma N}, g_{\omega N}$ and $g_{\rho N}$ as well as the parameters $b$  and $c$ are fitted to the bulk (binding energy, density, incompressibility, symmetry energy) and single particle (nucleon effective mass) properties of symmetric nuclear matter at saturation. They are taken here equal to those of the well known Glendenning--Moszkowski model (\cite{GM}). In particular, we consider the parametrizations GM1 and GM3 of this model to describe the pure nucleonic part of the system. The coupling of the $\Delta$ isobar with the different meson fields are poorly constrained due to the limited existence of experimental data. This leaves us with some freedom in the choice of these couplings. We consider two sets for the $\Delta$-meson couplings. The first one is the so-called universal coupling (UC) scheme  (\cite{Delta-strong1, Delta-strong2}) where all the $\Delta$-meson couplings are taken equal to those of the nucleons
\begin{equation}
 x_{\sigma \Delta} =1 \ , \,\, x_{\omega \Delta} =1 \ , \,\,  x_{\rho \Delta} = 1 \ ,
\end{equation}
where we have introduced the dimensionless couplings $x_{i\Delta}=\frac{g_{i\Delta}}{g_{iN}}$ ($i=\sigma, \omega,\rho$). The second set,  referred to from now on as stronger coupling (SC)  set, corresponds to the choice  $x_{\sigma \Delta} =1.15,
x_{\omega \Delta} =1,  x_{\rho \Delta} = 1$ . The interested reader is referred to  (\cite{delta1, delta2} )  for more details on this second set of parameters. 

Being the $d^*(2380)$ di-baryon an isospin singlet it couples only with the $\sigma$ and $\omega$ mesons. Unfortunately, there is currently no evidence that the interaction of $d^*(2380)$ with other particles is attractive or repulsive. In the lack of such information, in this work we explore a wide range of positive and negative values of the dimensionless couplings  $x_{id^*}=\frac{g_{id^*}}{g_{iN}}$ ($i=\sigma, \omega)$ in order to analyze the effect of both scenarios.

%%%%%%%%%%%%%%%%%%%%%%%%%%%%%%%%%%%%%%%%%%%%%%%%%%%%%%%%%%%%%%%%%%%%%%%%%%%%%%%%%%%%%%%%%%%%%%

\section{Results}\label{sec:res}

%%%%%%%%%%%%%%%%%%%%%%%%%%%%%%%%%%%%%%%%%%%%%%%%%%%%%%%
\begin{figure*}[t]
   \centering
\includegraphics[angle=0,width=0.7\textwidth]{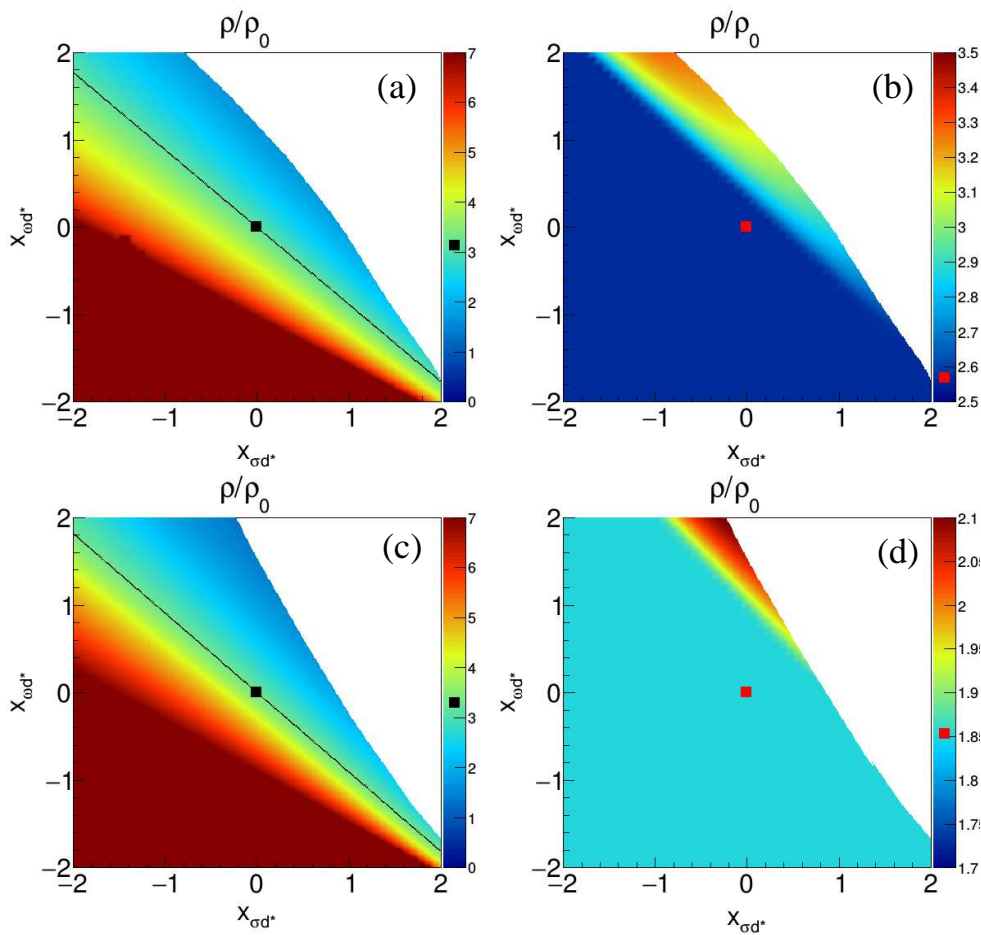}

   \caption{(Colors online) Onset density (in units of $\rho_0$) of the d$^*(2380)$ di-baryon(panels (a) and (c)) and the $\Delta^-$ isobar (panels (b) and (d)) in $\beta$-stable neutron star matter as a function of the dimensionless couplings $x_{\sigma d^*}$ and $x_{\omega d^*}$. Panels (a) and (b) correspond to the UC choice of parameters for the $\Delta$-meson couplings while panels (c) and (d) refer to the SC one. In all cases the nucleonic part is described with the GM1 parametrization of the Glendenning--Moszkowski model. The $d^*(2380)$ non-interacting case is highlighted with squares. The black line corresponds to solutions with where the onset density of the $d^*(2380)$ coincides with that of the non-interacting case.}
              \label{scandstarnoH}%
    \end{figure*}
%%%%%%%%%%%%%%%%%%%%%%%%%%%%%%%%%%%%%%%%%%%%%%%%%%%%%%

%%%%%%%%%%%%%%%%%%%%%%%%%%%%%%%%%%%%%%%%%%%%%%%%%%%%%%%
\begin{figure*}[t]
   \centering
                \includegraphics[angle=0,width=0.7\textwidth]{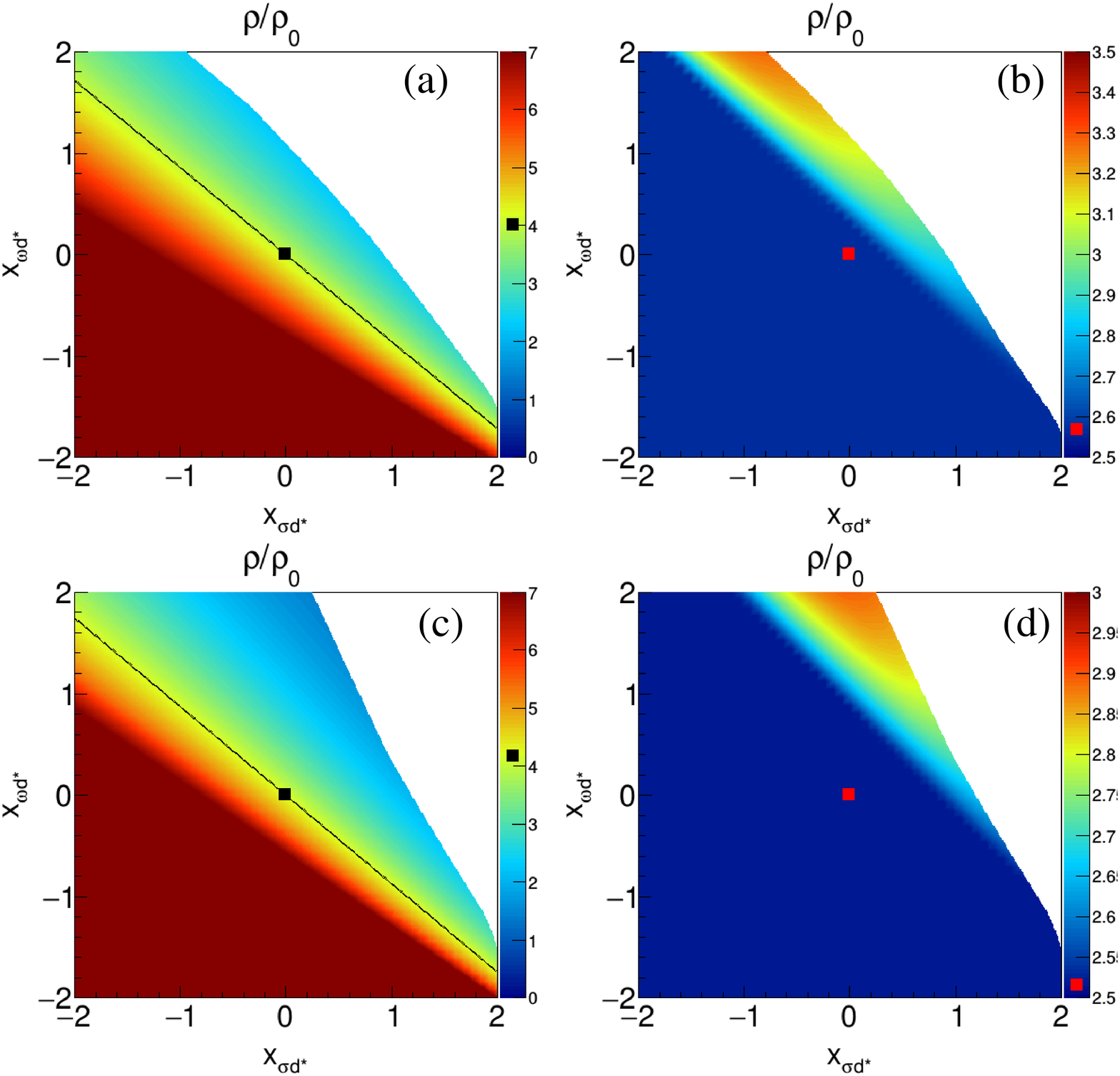}
   \caption{(Colors online) Same as Fig.\ \ref{scandstarnoH} with the nucleonic sector described using the GM3 model.}
              \label{scandstarnoH2}%
    \end{figure*}
%%%%%%%%%%%%%%%%%%%%%%%%%%%%%%%%%%%%%%%%%%%%%%%%%%%%%%

%%%%%%%%%%%%%%%%%%%%%%%%%%%%%%%%%%%%%%%%%%%%%%%%%%%%%%%
\begin{figure*}[t]
   \centering
               \includegraphics[angle=0,width=0.4\textwidth]{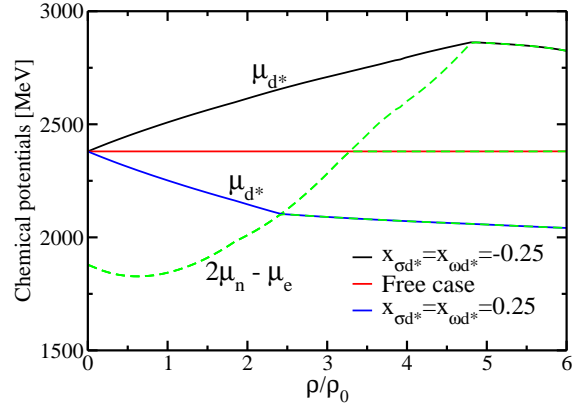}
      \caption{(Colors online) Chemical equilibrium condition for the appearance of the $d^*(2380)$ di-baryon in $\beta$-stable matter. Results are shown for the cases in which the $d^*(2380)$ feels attraction ($x_{\sigma d^*}=x_{\omega d^*}=0.25$), repulsion ($x_{\sigma d^*}=x_{\omega d^*}=-0.25$) or does not interact at all with the rest of the particles of the system. The GM1 parametrization together with the SC choice of the $\Delta$ couplings has been adopted.} 
              \label{cheq}%
    \end{figure*}
%%%%%%%%%%%%%%%%%%%%%%%%%%%%%%%%%%%%%%%%%%%%%%%%%%%%%%

%%%%%%%%%%%%%%%%%%%%%%%%%%%%%%%%%%%%%%%%%%%%%%%%%%%%%%%
\begin{figure*}[t]
\centering
\includegraphics[angle=0,width=0.7\textwidth]{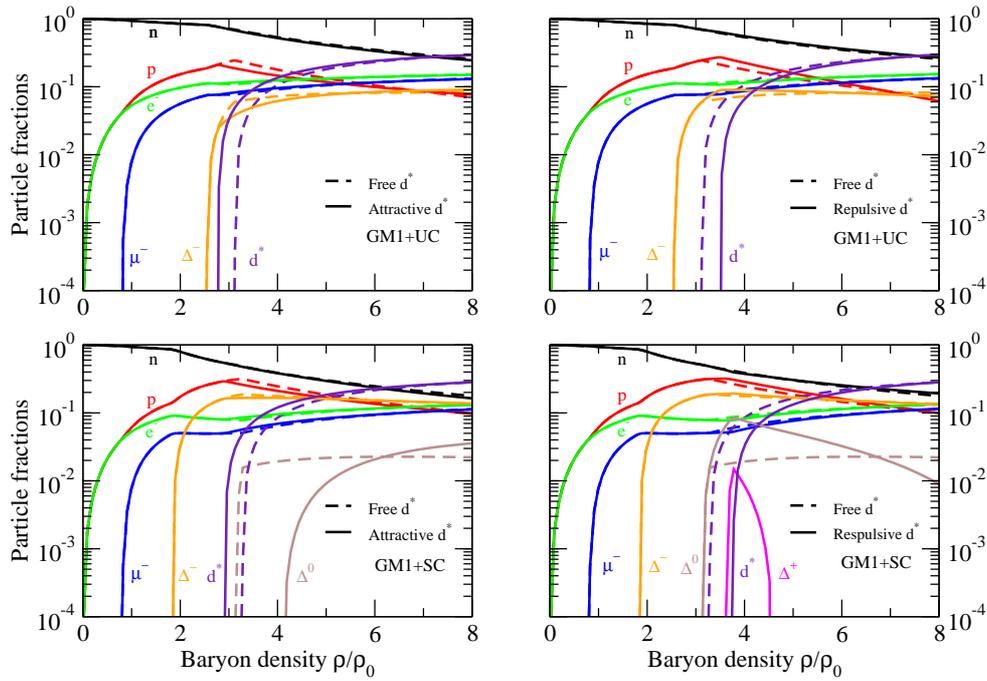}
\caption{(Colors online) Particle fraction as a function of the baryon density in units of $\rho_0$. The nucleonic part is described with the GM1 model. Results for the UC (SC) choice of parameters for the $\Delta$-meson couplings are shown in the upper (lower) panels. Left and right panels show results for a case in which the $d^*(2380)$ feels attraction ($x_{\sigma d^*}=x_{\omega d^*}=0.1$) or repulsion 
($x_{\sigma d^*}=x_{\omega d^*}=-0.1$), respectively. Dashed lines show the results for the free case for comparison.}
\label{chemic}%
\end{figure*}
%%%%%%%%%%%%%%%%%%%%%%%%%%%%%%%%%%%%%%%%%%%%%%%%%%%%%%

We start this section by showing in Fig.\  \ref{scandstarnoH} the onset density of the $d^*(2380)$ di-baryon (panels (a) and (c)) and the $\Delta^-$ isobar  (panels (b) and (d)) in $\beta$-stable neutron star matter as a function of  the dimensionless couplings $x_{\sigma d^*}$ and $x_{\omega d^*}$. It is worth noticing that, due its negative charge, the $\Delta^-$ is the first one of the four $\Delta$ resonances appearing in $\beta$-stable neutron star matter. Therefore, here we will focus only on it together with the $d^*(2380)$ di-baryon. The GM1 parametrization of the Glendenning--Moszkoski model have been used to describe the pure nucleonic part of the system. Results for the UC and SC choice of parameters for the $\Delta$-meson couplings are shown in panels (a,b) and (c,d), respectively. We notice that not all sets of values of $x_{\sigma d^*}$ and $x_{\omega d^*}$ lead to physical solutions. In particular, for some sets of couplings negative values of the pressure are obtained. These non-physical cases correspond to the blank regions in the four panels of the figure. We observe that the onset density of the $d^*(2380)$  varies significantly as a function of the couplings $x_{\sigma d^*}$ and $x_{\omega d^*}$, ranging from $\sim 2\rho_0$ up to $\sim 7\rho_0$. However, the onset of the $\Delta^-$ varies in a much smaller range, thus showing that in this particular scenario there is a very weak coupling between the two species. A similar conclusion holds using the GM3 parametrization instead of the GM1 one to describe the pure nucleonic part, as it is shown in Fig.\ \ref{scandstarnoH2}. Note, however, that in this case the SC choice of parameters for the $\Delta$-meson couplings leads to a later appearance of the $\Delta^-$, although this is not correlated with the appearance of the $d^*(2380)$. As a reference, we have also indicated the non-interacting $d^*(2380)$ case in both figures with solid squares. The black line, defined by the relation $x_{\omega d^*}=-0.88\,x_{\sigma d^*}$, shown in the left panel of both figures the case where a proper configuration of $x_{\sigma d^*}$ and $x_{\omega d^*}$ dimensionless couplings leads to an onset density of the $d^*(2380)$ equal to that of the non-interacting case. Note that below this line any combination of the $x_{\sigma d^*}$  and $x_{\omega d^*}$ couplings predicts an onset density of the $d^*(2380)$ larger than that of the free case. This an indication that the interaction of the $d^*(2380)$ is repulsive for all the values of these couplings in this region of the parameter space. Similarly, above the line $x_{\omega d^*}=-0.88\,x_{\sigma d^*}$ the onset density of $d^*(2380)$ predicted is always smaller than that of the non-interacting case and, consequently, the $d^*(2380)$ feels attraction for any value of the couplings $x_{\sigma d^*}$ and $x_{\omega d^*}$ sitting in this region. This is illustrated for the GM1 parametrization and the SC choice of the $\Delta$ couplings in Fig.\ \ref{cheq} where we show the chemical equilibrium condition for the appearance of the $d^*(2380)$ for the cases in which the $d^*(2380)$ feels attraction ($x_{\sigma d^*}=x_{\omega d^*}=0.25$), repulsion ($x_{\sigma d^*}=x_{\omega d^*}=-0.25$) or does not interact at all with the rest of the particles of the system. As seen in the plot  an attractive (repulsive) interaction  leads to a decrease (increase) of the $d^*(2380)$ chemical potential and, consequently, to an earlier (later) fulfillment of the equilibrium condition, $\mu_{d^*}=2\mu_n-\mu_e$, signaling the appearance of the $d^*(2380)$ in matter.

\begin{figure*}[t]
   \centering
               \includegraphics[angle=0,width=0.7\textwidth]{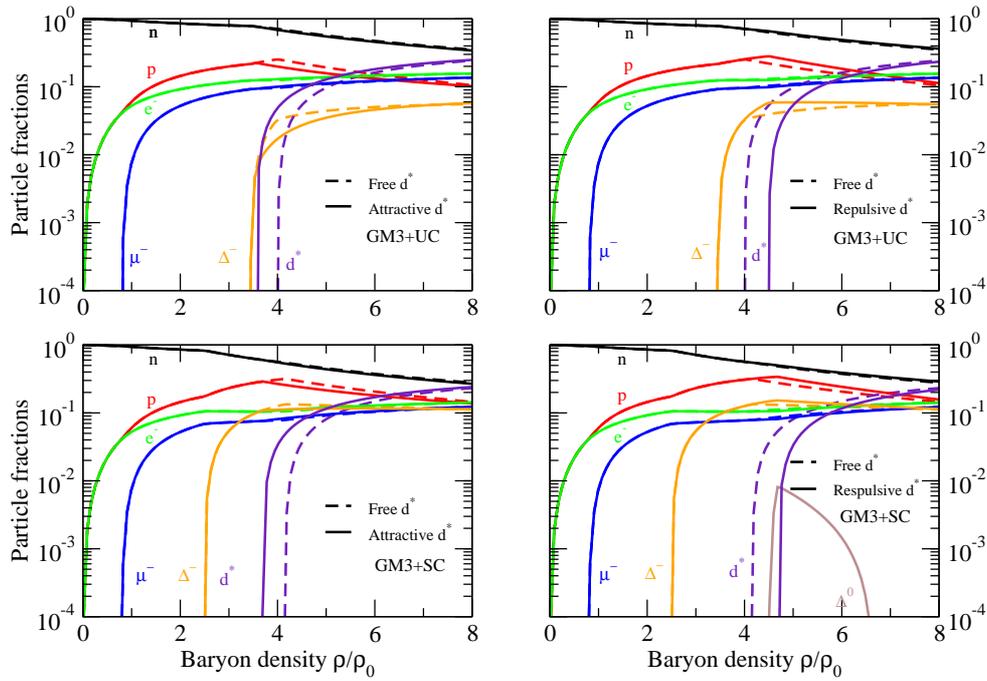}
      \caption{(Colors online) Same as Fig.\ \ref{chemic} with the nucleonic sector described using the GM3 model.} 
              \label{chemic2}%
    \end{figure*}

To better quantify the impact of the appearance of the $d^*(2380)$ in the medium, in Figs.\ \ref{chemic} and \ref{chemic2} we show the chemical composition of $\beta$-stable matter using the GM1 (Fig.\ \ref{chemic}) and GM3 (Fig.\ \ref{chemic2}) models to describe the pure nucleonic part and the UC (upper panels) or SC (lower panels) choice of parameters for the $\Delta$-meson couplings. The interaction of $d^*(2380)$ is assumed to be either attractive, with couplings $x_{\sigma d^*}=x_{\omega d^*}=0.1$ (left panels), or repulsive, with couplings $x_{\sigma d^*}=x_{\omega d^*}=-0.1$ (right panels). Results for the case in which the $d^*(2380)$ is assumed to be a free particle are also shown for comparison. Notice that the GM1 model leads in all the cases to an earlier appearance of the $\Delta$ isobar. Notice also, as mentioned before, that due to its negative charge the $\Delta^-$ appears at much lower densities than the other members of the $\Delta$ four-plet. Note in addition that with the UC choice for the $\Delta$-meson couplings, no other $\Delta$ resonances appear in the range of baryonic densities considered neither for the GM1 model nor for the GM3 one. On the contrary, when the SC set of couplings is adopted, also the $\Delta^0$ and $\Delta^+$ appear, both in the case of the GM1 model and only the former in the case of the GM3 one when the $d^*(2380)$ feels repulsion. The $\Delta^{++}$ is absent in all the cases. Finally, we also observe  that the onset density of the $\Delta^-$ is not affected by the attractive or repulsive character of the interaction of the $d^*(2380)$ with the other particles since it appears in all cases at lower densities than the $d^*(2380)$. The appearance of the $d^*(2380)$ induces an important and significant reduction of the neutron, proton and $\Delta$'s fractions since its baryon number is 2.  In addition, since the $d^*(2380)$ is positively charged, the lepton fractions increase in order to keep charge neutrality.

\begin{figure*}[t]
\centering
\includegraphics[angle=0,width=0.7\textwidth]{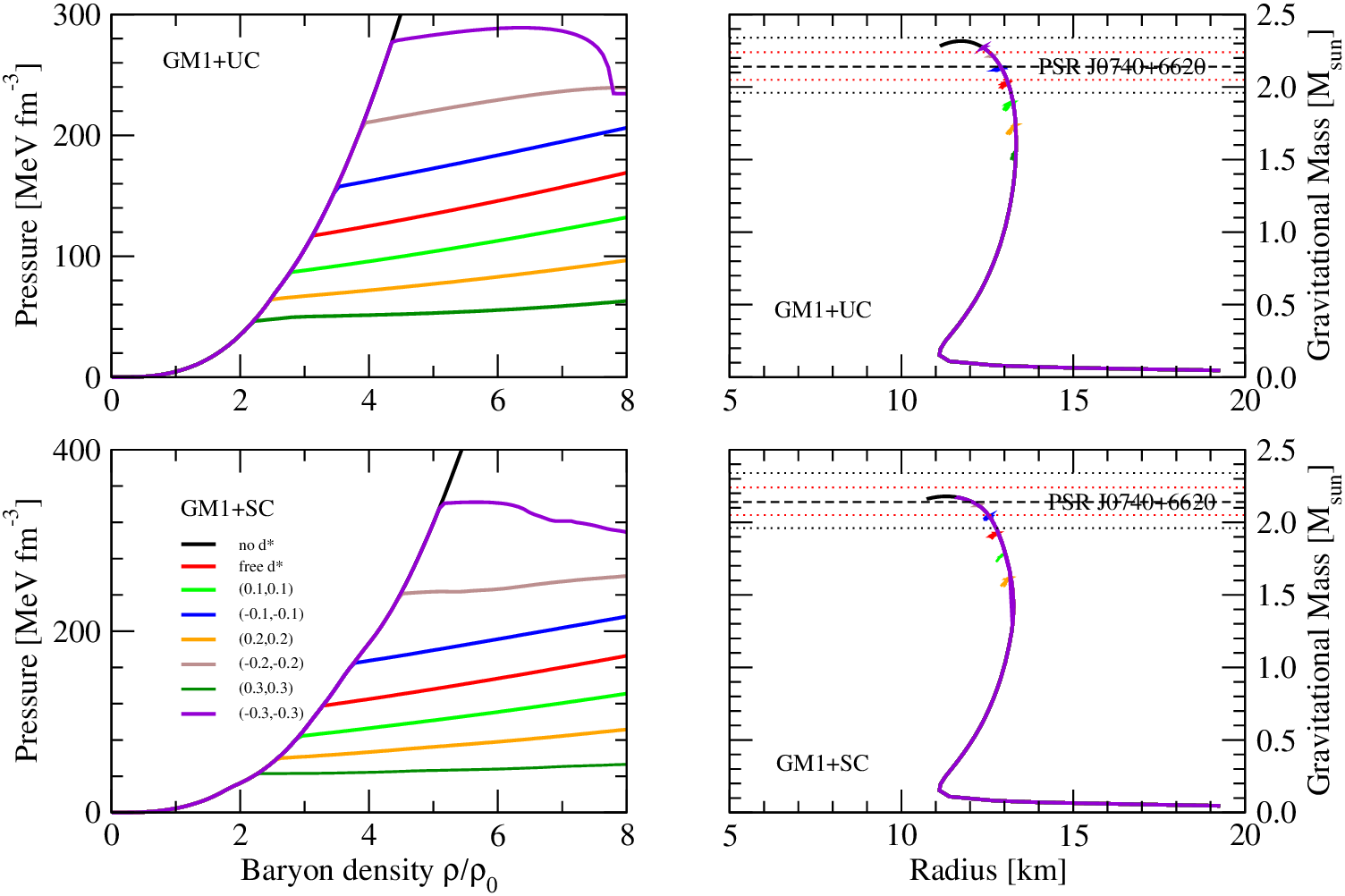}
\caption{(Colors online)Pressure (left panels) and mass-radius relation (right panels) obtained using the GM1 model to describe the pure nucleonic part of the EoS and the UC (upper panels) or the SC (lower panels) choice of parameters fro the $\Delta$-meson couplings. The interaction of $d^*(2380)$ is assumed to be either attractive, with couplings $(x_{\sigma d^*},x_{\omega d^*})=(0.1,0.1), (0.2,0.2)$ and $(0.3,0.3)$, or repulsive, with couplings $(x_{\sigma d^*},x_{\omega d^*})=(-0.1,-0.1), (-0.2,-0.2)$ and $(-0.3,-0.3)$. The recent measurement of the mass of the pulsar PSR J0740+6620 is shown with a $68.3\%$ credibility interval  ($2.14^{+0.10}_{-0.09}$ M$_\odot$) and $95.4\%$ credibility interval ($2.14^{+0.20}_{-0.18}$ M$_\odot$).}
\label{ns}%
\end{figure*}

\begin{figure*}[t]
\centering
\includegraphics[angle=0,width=0.7\textwidth]{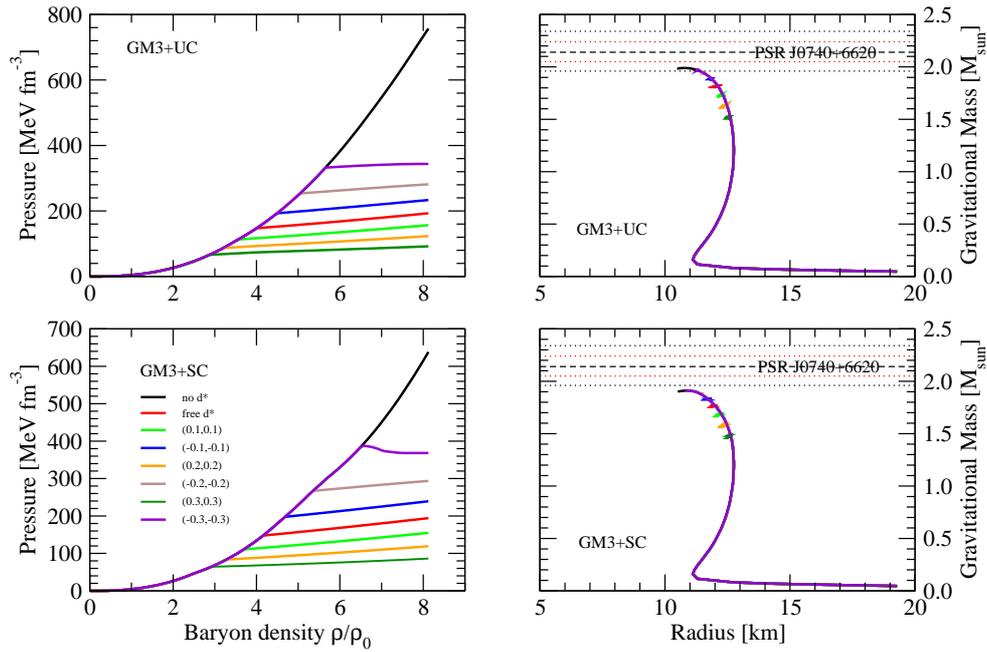}
\caption{(Colors online) Same as Fig.\ \ref{ns} with the nucleonic sector described using the GM3 model.}
\label{ns2}%
\end{figure*}

Let us now analyze the effect of the $d^*(2380)$ on the mass-radius relation of neutron stars and, in particular, on the maximum mass. To such end,  using our EoS together with that of Douchin and Haensel  (\cite{douchin2001unified}) for the low density stellar crust, we have solved the well known Tolmann--Oppenheimer--Volkoff (TOV) equations (\cite{tov1,tov2}) which describe the structure of non-rotating spherically symmetric stellar configurations in general relativity. In Figs.\ \ref{ns} and \ref{ns2}, we show the pressure (left panels) and the mass-radius relation (right panels) obtained using the GM1 (Fig.\ \ref{chemic}) and GM3 (Fig.\ \ref{chemic2}) models to describe the pure nucleonic part of the EoS and the UC (upper panels) or SC (lower panels) choice of parameters for the $\Delta$-meson couplings. The interaction of $d^*(2380)$ is assumed to be either attractive, with couplings $(x_{\sigma d^*},x_{\omega d^*})=(0.1,0.1), (0.2,0.2)$ and $(0.3,0.3)$, or repulsive, with couplings $(x_{\sigma d^*},x_{\omega d^*})=(-0.1,-0.1), (-0.2,-0.2)$ and $(-0.3,-0.3)$. Results for the case in with the presence of the $d^*(2380)$ is ignored and the case in which it is assumed to be a free particle are also shown for comparison. The first thing one observes when looking at the pressure of the system is that it is strongly reduced once the $d^*(2380)$ appears. This is simple due to: (i) the reduction of the neutron and proton fractions with the appearance of the $d^*(2380)$ with the consequent reduction of their partial contributions to the pressure, and (ii) the fact that the
$d^*(2380)$ itself does not contributes to the pressure. Note that the pressure continues increasing slowly after the appearance of the $d^*(2380)$ in the system except for the couplings $x_{\sigma d^*}=x_{\omega d^*}=-0.2$ and $x_{\sigma d^*}=x_{\omega d^*}=-0.3$ in the case of the GM1 model, and $x_{\sigma d^*}=x_{\omega d^*}=-0.3$ in the case of the GM3 one. In these cases the gradient of the pressure becomes negative at a given density. A negative gradient of the pressure is a signal for a mechanical instability which can give rise to a possible phase transition. Such a possibility, however, has not been considered in the present work. Therefore, those sets of $x_{\sigma d^*}$ and $x_{\omega d^*}$ couplings which lead to a negative gradient of the pressure at some given density represent solutions in which the appearance of the $d^*(2380)$ di-baryon induces the collapse of the neutron star into a black hole. 

Looking now into the mass-radius relation, one immediately notice that only the model GM1 (with both choices for the $\Delta$-meson couplings) predicts a maximum mass compatible with the recent measurement of the mass of the pulsar PSR J0740+6620 ($2.14^{+0.10}_{-0.09}$ M$_\odot$ ($2.14^{+0.20}_{-0.18}$ M$_\odot$) with a $68.3\%$ ($95.4\%$) credibility interval) (\cite{cromartie}) if the $d^*(2380)$ does not interact or feels slight repulsion. An attractive interaction of the $d^*(2380)$ leads to values of the maximum mass smaller than the highest one observed till now. Too much repulsion on the other side leads, as mentioned before, to a negative gradient of the pressure and, consequently, to the collapse of the star.

%We end this section by showing in Fig.\ \ref{MMnoH} the values for the maximum mass compatible with observation as a function of the dimensionless couplings $x_{\sigma d^*}$ and $x_{\omega d^*}$ for the two models GM1 (left panels) and GM3 (right panels) of the pure nucleonic part of the EoS and the UC (upper panels) and SC (lower panels) choices of the $\Delta$-meson couplings. Any other set of dimensional couplings outside the range shown in the figure leads, as already say, either to lower values of the maximum mass incompatible with the observed ones or to neutron star configurations that collapse into a black hole as soon as the $d^*(2380)$ appears.

%%%%%%%%%%%%%%%%%%%%%%%%%%%%%%%%%%%%%%%%%%%%%%%%%%%%%%%
%%%%%%%%%%%%%%%%%%%%%%%%%%%%%%%%%%%%%%%%%%%%%%%%%%%%%%

%%%%%

%%%%%%%%%%%%%%%%%%%%%%%%%%%%%%%%%%%%%%%%%%%%%%%%%%%%%%%
%%%%%%%%%%%%%%%%%%%%%%%%%%%%%%%%%%%%%%%%%%%%%%%%%%%%%%

%%%%%%%%%%%%%%%%%%%%%%%%%%%%%%%%%%%%%%%%%%%%%%%%%%%%%%%
%%%%%%%%%%%%%%%%%%%%%%%%%%%%%%%%%%%%%%%%%%%%%%%%%%%%%%

%%%%%%%%%%%%%%%%%%%%%%%%%%%%%%%%%%%%%%%%%%%%%%
    
\section{Conclusions}\label{sec:concl} 

This work represents an extension of a previous one (\cite{vid18}) where, assuming a simple free bosonic gas supplemented with a RMF model to describe the pure nucleonic part of the EoS,  we explored for the very first time the consequences that the presence of the $d^*(2380)$ di-baryon could have on the properties of neutron stars. Compared to that exploratory work, we have employed a standard non-linear Walecka model (\cite{dutrab}) within the framework of a relativistic mean field theory (RMF) including additional terms that describe the interaction of the $d^*(2380)$ di-baryon with the other particles of the system through the exchange of $\sigma$- and $\omega$-meson fields. The two well know parametrizations GM1 and GM3 of the Glendenning--Moszkowski model have been used to describe the pure nucleonic part of the EoS together with two different choices for the $\Delta$-meson couplings, namely the universal (UC) and the strong (SC) ones. Our results have showed that the presence of the $d^*(2380)$ leads to maximum masses compatible with the recent observations of $\sim 2$M$_\odot$ millisecond pulsars if the interaction of the $d^*(2380)$ is slightly repulsive or the $d^*(2380)$ does not interacts at all. An attractive interaction makes the EoS too soft to be able to support a $2$M$_\odot$ neutron star whereas an extremely repulsive one induces the collapse of the neutron star into a black hole as soon as the $d^*(2380)$ appears.
We conclude from our analysis that the presence of d$^*(2380)$ within a NS is plausible, although the exact density at which it appears and the production amount is still a matter for further scientific investigation.
Finally, it is worth mentioning, that the presence of $d^*(2380)$ particles in neutron star interior may induce new possible cooling mechanisms as discussed by (\cite{vid18}).

\begin{acknowledgements}

This work was partially supported by the STFC Grants No. ST/M006433/1 and ST/P003885/1, ST/L00478X/1 and by the COST Action CA16214 ``PHAROS: The multimessenger physics and astrophysics of neutron stars".
\end{acknowledgements}

%-------------------------------------------------------------------
\bibliographystyle{aa} % style aa.bst
\bibliography{biblio}
\end{document}